# Structural Distortion and Disappearance of Layer Ordering in Li$_x$CoO$_2 \cdot$2H$_2$O

Sangmoon Park[1], Yongjae Lee[1], and Thomas Vogt[1,2,*]

[1]Physics Department, Brookhaven National Laboratory, Upton, NY 11973-5000

[2]Center for Functional Nanomaterials, Brookhaven National Laboratory, Upton, NY 11973-5000

*The structure of a new member of the A$_x$CoO$_2$ x nH$_2$O family (A=Li,Na,K) was determined by high-resolution synchrotron X-ray powder diffraction. Layered lithium cobalt oxide hydrate Li$_x$CoO$_2$ x 2H$_2$O crystallizes in a disordered monoclinic structure (space group C2/m) with lattice parameters a = 4.8915(2) Å, b = 2.8239(1) Å, c = 10.7033(9) Å, $\beta$ = 112.386(4)°. Unlike the superconducting sodium cobalt oxide hydrate, this material possesses a structure with lithium-water layers intercalated between disordered octahedral sheets of cobalt oxide. The interlayer spacing is slightly larger (~0.9%) due to the higher water content, and one of the two lithium sites extends into the water layer. This material shows no superconducting transition above 2K.*

The discovery of superconductivity in the layered cobalt oxide hydrate Na$_{0.3}$CoO$_2$ x 1.4H$_2$O near 5K has marked an important milestone in the search for layered oxide superconductors without copper.[1] The layered cobalt oxide system is the first example of a superconductor where the presence of intercalated water is essential to providing the spatial separation needed for superconductivity. Further physical and chemical investigations of this new class of materials are being pursued. The chemical instability of many layered metal oxyhydrate system severely hampers synthesis efforts as well as proper physical characterizations.[2] In general, the dimensions of the interlayer spacings can be adjusted by the amount of materials intercalated and by chemical substitutions of the interlayer cations such as Na, K and Li, while the electrical and magnetic properties of the layer can be varied by substituting different transition metals into the layer or by changing the valence of the hexa-coordinated metals. We have been testing the adaptability and resulting structural and property changes in the general family of A$_x$MO$_2$ x nH$_2$O materials [3,4] and here we report the synthesis and structure of the novel lithium analogue of layered cobalt oxyhydrate Li$_x$CoO$_2$ x 2 H$_2$O.

Analog to our synthesis of superconducting Na$_{0.3}$CoO$_2$ x 1.4 H$_2$O Li$_x$CoO$_2 \cdot$2H$_2$O powder was prepared using Na$_2$S$_2$O$_8$ (Alfa 98%) as an oxidizing reagent. The precursor, LiCoO$_2$, was obtained by heating the mixture of Li$_2$CO$_3$ with a 10 mol% excess (Alfa 99.999%) and Co$_3$O$_4$ (Alfa 99.7%) at 900°C for 1d in air. LiCoO$_2$ with excess Na$_2$S$_2$O$_8$ (1:10) was placed in 20mL DI water. After adding 4 drops of 1N NH$_4$OH (pH ~10.5) in a beaker covered with a Parafilm$^{TM}$ the solution was stirred for one day, filtered, and dried in air. The so obtained sample was then further oxidized using excess Na$_2$S$_2$O$_8$. The sample was kept in a humidified container or a refrigerator and characterized by X-ray powder diffraction. We monitored the progress of the oxidation and intercalation process by x-ray diffraction: Figure 1 shows the x-ray diffraction pattern taken during the oxidation process of LiCoO$_2$ to Li$_x$CoO$_2 \cdot$yH$_2$O followed by a dehydration process where the c-axis collapses again. The X-ray pattern of the precursor of LiCoO$_2$ is shown in Figure 1 (b), respectively. During the oxidation process the *(0 0 l)* peaks shift indicating that the c axis expands from 13.9Å to 16.5 Å (R-3m setting) during the first oxidation process as shown in Figure 1 (c). After the second oxidation step two phases were detected as shown in Figure 1(d). A single phase of $Li_xCoO_2 \cdot 2H_2O$ was finally obtained by a third oxidation step (Figure 1 (e)). If no precautions are taken $Li_xCoO_2 \cdot yH_2O$ will dehydrate at ambient condition as shown in Figure 1 (f). High-resolution synchrotron X-ray powder diffraction data of $Li_xCoO_2 \cdot 2H_2O$ were measured at beam line X7A at National Synchrotron Light Source at Brookhaven National Laboratory. The sample was contained in a humidity chamber prior to the experiment (constant exposure to water vapor pressure at room temperature). A powder sample of $Li_xCoO_2 \cdot 2H_2O$ was quickly loaded into a 0.3mm glass capillary. The capillary was subsequently sealed and mounted on the 2nd axis of the diffractometer. A monochromatic beam of 0.7101(1) Å was selected using a channel-cut Ge(111) monochromator. A gas-propotional position-sensitive detector (PSD), gated at the Kr-escape peak, was employed for high-resolution ($\Delta d/d \approx 10^{-3}$) powder diffraction data measurements.[5] The PSD was stepped in 0.25° intervals between 3° and 65° with an increasing counting time at higher angle. The capillary was spun during the measurement to provide better powder averaging. The first three peaks in the resulting powder pattern are indicative of the interlayer spacing similar to what was observed in the sodium cobalt oxyhydrate.[1] However, the higher angle peaks could not be indexed in a hexagonal unit cell stting. Subsequent indexing of the powder pattern led to a monoclinic cell with a *c* axis length close to *c*/2 of the hexagonal sodium analogue. The structure was solved *ab initio* assuming a layered structure with cobalt atoms (Co1) at (0,0,0). All refinements were carried out using the Rietveld method[6,7]. Difference Fourier maps were generated and allowed us to locate octahedrally coordinating oxygen atoms (O1) at (~0.6,0,~0.9). The resulting profile fit was, however, rather poor. Successive difference Fourier sections were calculated to locate other intercalated species between the layers. Subsequently extra electron density was found at the (0.5,0,0) site within the octahedral layer. This site was then modeled using another cobalt atom (Co2) along with a coordinating oxygen atom (O2) at (~0.1,0,~0.9) site. Using this model, the fit improved by ~50%. In the final refinement, restraints were used for the isotropic displacement parameters for the atoms within the octahedral layers and those intercalated between the layers. Due to their small x-ray scattering factor the occupancies for lithium atoms were fixed to unity despite the fact that during the oxidation process Li is deintercalated and Co oxidized. Asymmetry terms of Finger, Cox and Jephcoat's formalism[8,9] were then introduced to model diffraction peak shapes, and a March-Dollase ellipsoid model was used to account for preferred orientations[10]. The final profile fit is depicted in Figure 2. The refined overall structural model is summarized in Table 1 and selected interatomic distances are listed in Table 2.

The structure of $Li_xCoO_2 \cdot 2H_2O$ consists of disordered layers of cobalt oxide octahedra with water molecules and lithium cations intercalated in-between (Fig. 3). This is an example of a layered compound showing a complete statistical disordering in both inter- and intra-layer sites. As a result of the monoclinic distortion, two sets of Co-O distances are found in each octahedral layer with two long bonds running along the *a*-axis. Our model shows two intercalated lithium layers stacked perpendicular to the *c*-axis, which contrasts the single layer distribution of the sodium in the hexagonal sodium cobalt oxyhydrate. Although the model shows a fixed full occupancy of the two lithium sites leading to one lithium atoms per formula unit, it does not represent the true amount intercalated X-ray data are not sensitive to allow refinement of the occupancy of such a light element. We were not able to dissolve our sample and perform elemental analysis and therefore denote the stoichiometry as $Li_x$.  It is interesting to note that the deintercalation of lithium and the insertion of $H_2O$ into $LiCoO_2$ leads to a monoclinic distortion of the parent rhombohedral lattice [11].  A monoclinic distortion was also observed in the pressure-induced hydration of the superconducting $Na_{0.3}CoO_2·1.4H_2O$.[3] The crystallographically refined water content of 2.0 per formula unit matches the one determined from thermal gravimetric analysis (TGA) (27 wt.% loss up to 250 °C, Fig. 4). The comparison of the thermal decomposition of $Na_{0.3}CoO_2·1.4H_2O$ and $Li_xCoO_2·2H_2O$ in Figure 4 shows the latter to be initially more stable. The higher water content of $Li_xCoO_2$ x $2H_2O$ compared to the sodium cobalt oxyhydrate analogue results in a slight expansion of the interlayer spacing (by ~0.9%) despite the smaller size of the Li cation.. It appears that a second hydrate phase $Li_xCoO_2·0.3H_2O$ exists in analogy to $Na_{0.3}CoO_2$ x $0.6H_2O$.. The water molecules in $Li_xCoO_2·2H_2O$ hydrogen bond with each other and are coordinated to the lithium cations.  As observed in the hexagonal sodium analogue, the water content in the lithium cobalt oxyhydrate is also very sensitive to humidity and temperature near ambient conditions, and exposure to laboratory humidity (typically ~25%) results in a sequential loss of intercalated water molecules.  A SQUID magnetometer (Quantum Design) was used to determine the magnetic susceptibility of the lithium oxyhydrate as a function of temperature.  We observed no diamagnetism indicative of superconductivity and X-ray diffraction experiment down to 15 K also suggest that there are no low temperature phase transitions.

The distinct structural features of the lithium cobalt oxyhydrate with its disordered metal oxide layers present a new basis to probe the structural origin of superconductivity in the $A_xCoO_2$ x $nH_2O$ family of layered cobalt oxide hydrate systems. We plan to investigate possible solid-solution series as well as the structure and magnetization of these systems under pressure.

This work was supported by an LDRD from BNL.  The authors thank K. Kang and L. Lewis for access to TGA measurement facilities and C. Petrovic for the magnetization measurement.  Research carried out in part at the NSLS at BNL is supported by the U.S. DOE (DE-Ac02-98CH10886 for beam line X7A).

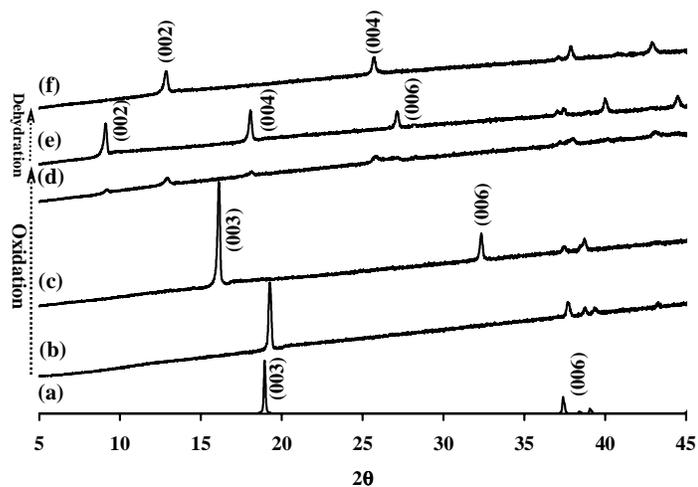

Figure 1. (a) Simulated X-ray pattern (Cu Kα radiation) for LiCoO$_2$ (ICSD #92183). Experimental powder X-ray diffraction patterns for Li$_x$CoO$_2$·yH$_2$O samples taken during the oxidation and intercalation process using Na$_2$S$_2$O$_8$ shows the expansion of the c-axis [(b) to (e)]. The dehydration observed at ambient conditions when no precautions are taken leads to the collapse of the c-axis [(f)].

Table 1. Atomic coordinates of $Li_xCoO_2 \cdot 2H_2O$.

| atom | x | y | z | occu. | $U_{iso}$ (Å$^2$) |
|---|---|---|---|---|---|
| Co1 | 0.000 | 0.000 | 0.000 | 0.5 | 0.0054(4) |
| Co2 | 0.500 | 0.000 | 0.000 | 0.5 | 0.0054(4) |
| O1 | 0.619(6) | 0.000 | 0.904(4) | 0.5 | 0.0054(4) |
| O2 | 0.103(5) | 0.000 | 0.916(4) | 0.5 | 0.0054(4) |
| Li1 | 0.10(4) | 0.000 | 0.37(4) | 0.25 | 0.168(11) |
| Li2 | 0.5000 | 0.000 | 0.500 | 0.5 | 0.168(11) |
| OW1 | 0.69(2) | 0.000 | 0.638(3) | 0.5 | 0.168(11) |
| OW2 | 0.04(1) | 0.000 | 0.380(5) | 0.5 | 0.168(11) |

Space group $C2/m$; a = 4.8915(2) Å, b = 2.8239(1) Å, c = 10.7033(9) Å, β = 112.386(4)°, V = 136.70(2) Å$^3$. $_wR_p$ = 4.52%, $R_p$ = 2.53%. OW denotes water molecule modeled with oxygen atom. Restraints were used to tie isotropic displacement parameters same for layer and interlayer species, respectively.

Table 2. Selected interatomic distances (Å) of $Li_xCoO_2 \cdot 2H_2O$.

| Co1-O1 | 1.752(29) × 2 | Co2-O2 | 1.802(27) × 2 |
|---|---|---|---|
|  | 1.961(18) × 4 |  | 1.842(17) × 4 |
| Li1-O2 | 2.8(4) |  |  |
| Li1-OW1 | 2.01(13) | Li2-OW1 | 2.84(5) |
|  | 3.00(8) |  |  |
| Li1-OW2 | 2.52(18) | Li2-OW2 | 1.96(3) |
|  | 2.85(4) |  | 2.13(6) |
|  | 2.94(24) |  |  |
| OW1-O2 | 2.89(5) | OW2-O2 | 2.98(7) |
| OW1-OW1 | 2.8239(1) | OW2-OW2 | 2.73(10) |
|  | 2.83(7) |  | 2.8239(1) |
|  | 2.8240(1) |  | 2.8240(1) |
| OW1-OW2 | 2.93(5) |  | 2.97(11) |

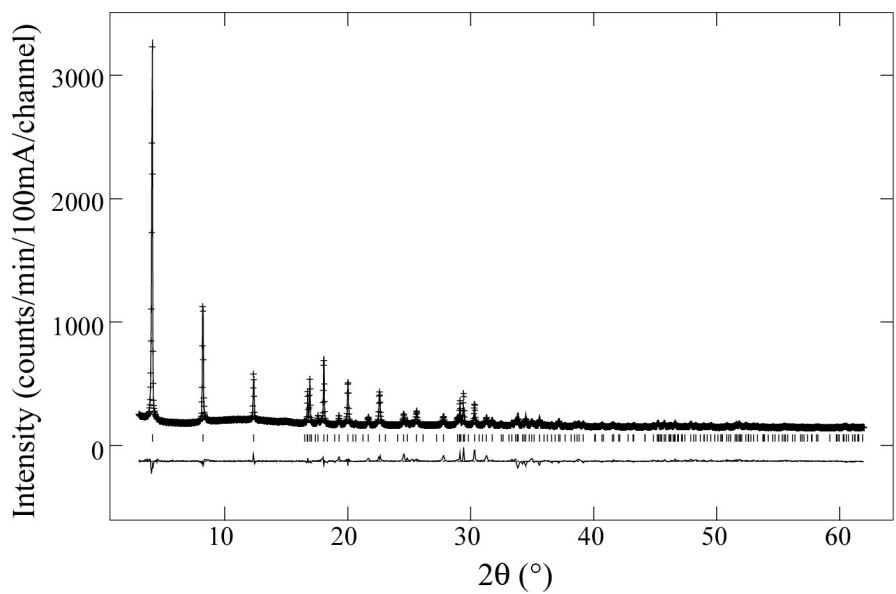

Figure 2.     Rietveld refinement plot of the structural model of $Li_xCoO_2·2H_2O$ at room temperature compared to synchrotron X-ray powder diffraction data.  Points shown represent observed data.  The continuous lines through the sets of points are the calculated profiles from the structure refinements summarized in Tables.  Fixed points were used to model the structured backgrounds.  The sets of tick marks below the data indicate the positions of the allowed reflections.  The lower curves represent the differences between observed and calculated profiles ($I_{obs} - I_{calc}$) plotted on the same scale as the observed data.

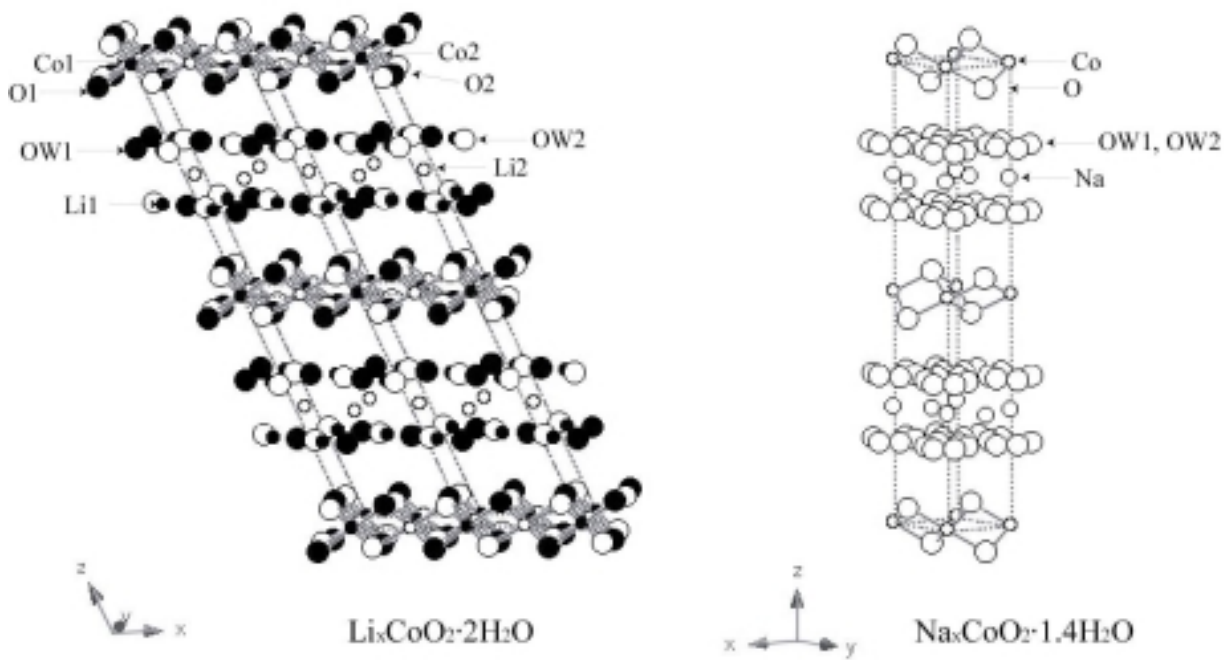

Figure 3.    A ball and stick representation of the structure of $Li_xCoO_2·2H_2O$ (left) viewed perpendicular to the c-axis. For comparison, the structure of $Na_{0.3}CoO2·1.4H_2O$ is shown in the right. The disordering of $Li_xCoO_2·2H_2O$ is depicted with sets of closed and open circles to represent statistical distributions of layer and interlayer atoms. Dotted lines in the figures define monoclinic (×4) and hexagonal unit cell of $Li_xCoO_2·2H_2O$ and $Na_xCoO2·1.4H_2O$, respectively.

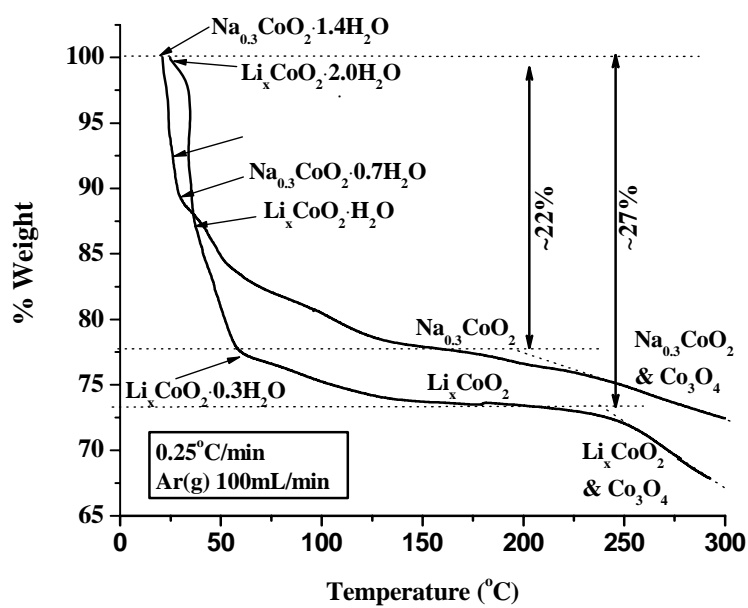

Figure 4. Thermogravimetric analysis (TGA) of $Na_{0.3}CoO_2 \cdot 1.4H_2O$ and $Li_xCoO_2 \cdot 2H_2O$ with heating rate $0.25^\circ C/min$ under Ar (g) flowing.

# References

[1] K. Takada, H. Sakirai, E. Takayama-Muromachi, *et al*., Nature **422**, 53 (2003).

[2] M. L. Foo, R. E. Schaak, V. L. Miller, *et al*., Solid State Commun. **127**, 33 (2003).

[3] S. Park, Y. Lee, A. Moodenbaugh, *et al*., Phys. Rev. B **68**, 180505(R) (2003).

[4] S. Park, T. Vogt, Y. Lee, unpublished

[5] G. C. Smith, Synch. Rad. News **4**, 24 (1991).

[6] H. M. Rietveld, J. Appl. Crystallog. **2**, 65 (1969).

[7] B. H. Toby, J. Appl. Crystallogr. **34**, 210 (2001).

[8] P. Thompson, D. E. Cox, and J. B. Hastings, J. Appl. Crystallogr. **20**, 79 (1987).

[9] L. W. Finger, D. E. Cox, and A. P. Jephcoat, J. Appl. Crystallogr. **27**, 892 (1994).

[10] W. A. Dollase, J. Appl. Crystallogr. **19**, 267 (1986).

[11] T. Ohzuku and A. Ueda, J. Electrochem. Soc. **141**, 2972 (1994)